\documentclass[%
superscriptaddress,
preprint,
nofootinbib,
nobibnotes,
bibnotes,
 amsmath,amssymb,
 aps,
]{revtex4-1}

\usepackage{graphicx}% Include figure files
\usepackage{epsfig}% Include figure files
\usepackage{dcolumn}% Align table columns on decimal point
\usepackage{bm}% bold math
\usepackage[mathlines]{lineno}% Enable numbering of text and display math
\begin{document}
\raggedright{Prepared for submission on Physical Review D}
%\title{Temperature-related variations of the underground flux of atmospheric muons \\
%with 24 years of data of the Large Volume Detector}% Force line breaks with \\
\title{Characterization of the varying flux of atmospheric muons \\
measured with the Large Volume Detector for 24 years}

\author{N.Yu.~Agafonova}
\affiliation{Institute for Nuclear Research, Russian Academy of Sciences, Moscow, Russia}
\author{M.~Aglietta}
\affiliation {University of Torino and INFN-Torino, Italy}
\affiliation {INAF, Osservatorio Astrofisico di Torino, Italy}
\author{P.~Antonioli}
\affiliation {University of Bologna and INFN-Bologna, Italy}
\author{V.V.~Ashikhmin}
\affiliation{Institute for Nuclear Research, Russian Academy of Sciences, Moscow, Russia}
\author{G.~Bari}
\affiliation {University of Bologna and INFN-Bologna, Italy}
%\author{R.~Bertoni}
%\affiliation {University of Torino and INFN-Torino, Italy}
%\author{E.~Bressan}
%\affiliation {University of Bologna and INFN-Bologna, Italy}
%\affiliation {Centro Enrico Fermi, 00184 Roma, Italy}
\author{G.~Bruno}
\affiliation {INFN, Laboratori Nazionali del Gran Sasso, Assergi, L'Aquila, Italy}
\affiliation {New York University Abu Dhabi, NYUAD, United Arab Emirates}
\author{E.A.~Dobrynina}
\affiliation{Institute for Nuclear Research, Russian Academy of Sciences, Moscow, Russia}
\author{R.I.~Enikeev}
\affiliation{Institute for Nuclear Research, Russian Academy of Sciences, Moscow, Russia}
\author{W.~Fulgione}
\affiliation {INFN, Laboratori Nazionali del Gran Sasso, Assergi, L'Aquila, Italy}
\affiliation {INAF, Osservatorio Astrofisico di Torino, Italy}
\author{P.~Galeotti}
\affiliation {University of Torino and INFN-Torino, Italy}
\affiliation {INAF, Osservatorio Astrofisico di Torino, Italy}
\author{M.~Garbini}
\affiliation {University of Bologna and INFN-Bologna, Italy}
\affiliation {Centro Enrico Fermi, 00184 Roma, Italy}
\author{P.~L.~Ghia}
\affiliation{Institut de Physique Nucleaire, CNRS, 91406 Orsay, France}
\author{P.~Giusti}
\affiliation {University of Bologna and INFN-Bologna, Italy}
%\author{F.~Gomez}
%\affiliation {University of Torino and INFN-Torino, Italy}
%\affiliation {INAF, Osservatorio Astrofisico di Torino, Via Osservatorio, 30, I-10025 Pino Torinese (Torino), Italy\label{addr3}}
\author{E.~Kemp}
\affiliation {University of Campinas, Campinas, Brazil}
\author{A.S.~Malgin}
\affiliation{Institute for Nuclear Research, Russian Academy of Sciences, Moscow, Russia}
%\author{B.~Miguez}
%\affiliation {University of Campinas, Campinas, Brazil}
\author{A.~Molinario}
\affiliation {INFN, Laboratori Nazionali del Gran Sasso, Assergi, L'Aquila, Italy}
\affiliation {Gran Sasso Science Institute, L'Aquila, Italy}
\author{R.~Persiani}
\affiliation {University of Bologna and INFN-Bologna, Italy}
\author{I.A.~Pless}
\affiliation{Massachusetts Institute of Technology, Cambridge, USA}
%\author{A.~Porta}
%\affiliation {University of Torino and INFN-Torino, Italy}
%\affiliation {INAF, Osservatorio Astrofisico di Torino, Italy}
\author{S.Rubinetti}
\affiliation {University of Torino and INFN-Torino, Italy}
\affiliation {INAF, Osservatorio Astrofisico di Torino, Italy}
%\author{A.~Razeto}
%\affiliation {INFN,Laboratori Nazionali del Gran Sasso,Assergi, Italy}
%\author{V.G.~Ryasny}
%\affiliation{Institute for Nuclear Research, Russian Academy of Sciences, Moscow, Russia}
\author{O.G.~Ryazhskaya}
\affiliation{Institute for Nuclear Research, Russian Academy of Sciences, Moscow, Russia}
%\author{O.~Saavedra}
%\affiliation {INFN-Torino, OATO-Torino and University of Torino, Italy}
\author{G.~Sartorelli}
\affiliation {University of Bologna and INFN-Bologna, Italy}
\author{I.R.~Shakiryanova}
\affiliation{Institute for Nuclear Research, Russian Academy of Sciences, Moscow, Russia}
\author{M.~Selvi}
\affiliation {University of Bologna and INFN-Bologna, Italy}
\author{C.Taricco}
\affiliation {University of Torino and INFN-Torino, Italy}
\author{G.~C.~Trinchero}
\affiliation {University of Torino and INFN-Torino, Italy}
\affiliation {INAF, Osservatorio Astrofisico di Torino, Italy}
\author{C.~F.~Vigorito\footnote{Corresponding author: C.F. Vigorito, Dipartimento di Fisica, Universit\`a di Torino, Italy email: vigorito@to.infn.it tel. +39 011 6707349}}
\affiliation {University of Torino and INFN-Torino, Italy}
\author{V.F.~Yakushev}
\affiliation{Institute for Nuclear Research, Russian Academy of Sciences, Moscow, Russia}
\author{A.~Zichichi}
\affiliation {University of Bologna and INFN-Bologna, Italy}
\affiliation {Centro Enrico Fermi, 00184 Roma, Italy}

\collaboration{LVD Collaboration}
\noaffiliation

\begin{abstract}
The Large Volume Detector (LVD), hosted in the INFN Laboratori Nazionali del Gran Sasso, is triggered by atmospheric muons at a rate of  $\sim 0.1$~Hz. 
The data collected over almost a quarter of century are used to study the muon intensity underground. 
The 50-million muon series, the longest ever exploited by an underground instrument, allows for the accurate long-term monitoring of the muon intensity underground. 
This is relevant as a study of the background in the Gran Sasso Laboratory, which hosts a variety of long-duration, low-background detectors. 
We describe the procedure to select muon-like events as well as the method used to compute the exposure. 
We report the value of the average muon flux measured from 1994 to 2017: $\mathrm{I_{\mu}^0 = 3.35 \pm 0.0005^{stat}\pm 0.03^{sys} \cdot 10^{-4} ~m^{-2} s^{-1}}$. 
We show that the intensity is modulated around this average value due to temperature variations in the stratosphere. We quantify such a correlation by using temperature data from the European Center for Medium-range Weather Forecasts: we find an effective temperature coefficient $\mathrm{\alpha_{T}} = 0.94\pm0.01^{stat} \pm0.01^{sys}$, in agreement with other measurements at the same depth.
%This is the most precise measurement underground at a minimal depth of 3000 m w.e..
We scrutinise the spectral content of the time series of the muon intensity by means of the Lomb-Scargle analysis. This yields the evidence of a 1-year periodicity, as well as the indication of others, both shorter and longer, suggesting that the series is not a pure sinusoidal wave. Consequently, and for the first time, we characterise the observed modulation in terms of amplitude and position of maximum and minimum on a year-by-year basis.

\end{abstract}

\pacs{Valid PACS appear here}% PACS, the Physics and Astronomy

\maketitle

%\tableofcontents

\section{Introduction}
\label{intro}

When high-energy cosmic rays enter the atmosphere, they produce a large number of secondary particles, in a series of successive interactions with atmospheric nuclei, called extensive air showers (EAS). EAS particles produced in the upper atmosphere propagate longitudinally through the atmosphere: at ground level, the most abundant among them are muons, which are produced in the decay of short-lived mesons, namely charged pions and kaons. Thanks to their small energy loss, small cross-section and long lifetime, higher energy (above $\sim$ 1 TeV) muons can penetrate deeply underground. Thus, large acceptance instruments located underground, originally designed for, e.g., neutrino or proton decay studies, all have excellent capabilities for the study of high energy atmospheric muons. The Large Volume Detector (LVD) \cite{1992Aglietta}, located in the INFN Laboratori Nazionali del Gran Sasso (LNGS) at a minimal depth of 3100 m w.e., is one of such detectors. Despite the large amount of overhead rock, LVD is triggered by atmospheric muons at a rate of $\sim 0.1$ Hz. 

Underground muons are exploited for a variety of physics analyses, most notably for the measurement of the flux and the composition of Galactic cosmic rays (see e.g.,\cite{1990Gaisser} and \cite{1993Lipari}) as well as for the search for anisotropies (see \cite{1997Munakata}, \cite{2003Macro} and \cite{2007SK}). Also, the study of muons underground allows for the measurement of the high-energy part (above 1 TeV) of the sea-level muon energy spectrum through the depth-intensity relation (e.g., \cite{1998Aglietta}). Finally, the steady flux of muons underground is used for the validation and calibration of deep detectors. However, cosmic-ray muons are also one of the unavoidable backgrounds in underground laboratories for experiments searching for rare events. While muons can be rejected quite efficiently through either dedicated vetoes or selection criteria, more irksome is the background due to fast neutrons produced by their interactions in the rock. Although the rate of muon-induced neutrons is more than one order of magnitude smaller than that of radiogenic neutrons, the former have a much harder energy spectrum, extending to several GeV. Not only they can easily penetrate the detectors shielding, but also they can interact and generate secondary neutrons in the MeV range. Muon-induced neutrons can thus mimic events in underground detectors such as those looking for dark matter, double-beta decay or studying neutrino properties and sources (see for example \cite{2018Baudis}, \cite{2016Vissani}, \cite{2018Lisi} and references therein).

The study of the muon flux underground, whose intensity depends on the specific site, is thus relevant to characterise one of the most important background for deep detectors. It is the objective of this work, which exploits data collected with LVD over almost a quarter of century. Such a long-term measurement allows us to characterise the variations of the flux, important in view of long-duration instruments looking for rare events. Indeed, it has been known since the 1950s \cite{1952Barret}\cite{1954Barrett} that the intensity of atmospheric muons is affected by the temperature in the stratosphere. The parent mesons either interact again and produce further cascades of secondaries, or decay into muons. If the temperature gets higher, the air density gets lower: this reduces the probability of meson interaction, in turn yielding, 
for pions or kaons in a different way, a larger fraction decaying to produce muons, resulting in a higher muon rate. The temperature of the stratosphere, although more stable than that of the troposphere, is subject to variations with different periods. The seasonal modulation is the dominant one, although its amplitude can be modulated by other secondary variations, such as those due to the so-called Sudden Stratospheric Warming (SSW) events \cite{1952Schrhag}, or to the 11-years solar cycle (see e.g. \cite{Randel2009},\cite{2015Kuchar}).

The annual cycle induces an annual variation on the muon flux measured underground, observed by several detectors 
\cite{1954Barrett}, \cite{1954Sherman}, \cite{1981Utah}, \cite{1991Baksan}, \cite{1999Amanda}, \cite{2003Macro},  \cite{2011Icecube}, \cite{2014Minos}, \cite{2016Agostini}, \cite{2017Opera}, \cite{2017DoubleChooz}, \cite{2018DayaBay}, \cite{2019Agostini}.
In an earlier investigation \cite{2009Selvi}, we measured such modulation using 8 years of muons detected by LVD. In this work we improve and update the previous study by using data detected with a three times larger exposure, corresponding to a time-series of 24 years from 1994 to 2017, the longest ever exploited by an underground instrument. Thanks to the large accumulated number of events, we are able in this work to measure with high precision the coefficient of correlation between the muon intensity and the temperature. Moreover, the large statistics allows us, for the first time, to characterise in terms of amplitude and position of the maximum the annual modulation of the muon intensity on a year-by-year basis.

%The measurement of the muon intensity underground, and of its variations, is of interest because muons (and secondaries produced in their interactions, such as neutrons) represent one of the unavoidable backgrounds in underground laboratories. In particular, the long-term monitoring of such background is relevant in view of long duration instruments looking for dark-matter signals via the annual-modulation signature \cite{2018Bernabei}.
%The study of the time-series of more than fifty millions of muons collected by LVD in 24 years in relation with the temperature in the stratosphere is the focus of this paper, which is organized as follows.
The paper is organised as follows.
In Section \ref{sec.T} we introduce the concept of effective temperature and describe how this is calculated starting from data of the European Center for Medium-range Weather Forecasts (ECMWF) \cite{2011Dee}. 
In section \ref{sec.mu} we describe the LVD detector, we detail the criteria used to select the muon data set used in this work and we calculate the muon flux as a function of time.
In Section \ref{corr} we extract the correlation coefficient from the analysis of the variations of the muon flux associated to those of the temperature.
Finally, in Section \ref{spect} we perform a spectral analysis of the muon and temperature time series and we determine the amplitude and position of the maximum of the modulation on a year-by-year basis.
Discussion and conclusions in Section \ref{sum}.

\section{The temperature data set}
\label{sec.T}

For the purpose of this analysis we exploit the temperature profile of the atmosphere provided by the European Center for Medium-range Weather Forecasts \cite{2011Dee}, for the time period Jan.1st, 1994 - Dec.31st, 2017.
%http://badc.nerc.ac.uk/$\sim$view/adc.nerc.ac.uk ATOM dataent 12458543158227759 
It is compiled on the basis of different types of observations (e.g., surface, satellite, and upper air sounding) at many locations; a global atmospheric model is then used to interpolate it to a particular location. As for the latter, we consider in this analysis the coordinates of the LNGS: 13.5333$^{\circ}$ E, 42.4275$^{\circ}$ N. The model provides atmospheric temperatures at 37 discrete pressure levels in the [1-1000] hPa range, four times a day, namely at 00.00 h, 06.00 h, 12.00 h and 18.00 h UTC.

To study the impact of the temperature on the number of recorded muons, we need to account for the fact that the atmosphere is non-isothermal: variations occur differently at different pressure levels. This is done by combining the temperatures at each level into a unique ``effective'' temperature, $\mathrm{T_{eff}}$, as introduced by \cite {1952Barret} and developed in \cite{1997MACRO} and \cite{2010Minos} and references therein. 
In short, the effective temperature is a weighted average over several altitudes, the weight being larger for altitudes at which the air density is lower and hence mesons more probably decay into muons. Namely, to calculate $\mathrm{T_{eff}}$ we use:

\begin{equation}\label{teff}
\mathrm{T_{eff} = \,\frac{\Sigma_{n=1}^{N}\,\Delta X_{n} T(X_{n})W(X_{n})}{\Sigma_{n=1}^{N}\,\Delta X_{n} W(X_{n})} }
\end{equation}

where N=37 is the number of pressure levels at which temperature is available, $\mathrm{T(X_n)}$ is the temperature at the atmospheric depth $\mathrm{X_n}$, $\mathrm{\Delta X_{n}}$ is the thickness at the depth $\mathrm{X_n}$, varying between 1 and 25 hPa depending on the altitude and $\mathrm{W(X_{n})}$ is the weight at $\mathrm{X_n}$.

\begin{figure*}[htbp]
\centering
\vspace{.5cm}
\includegraphics[width=9cm,clip]{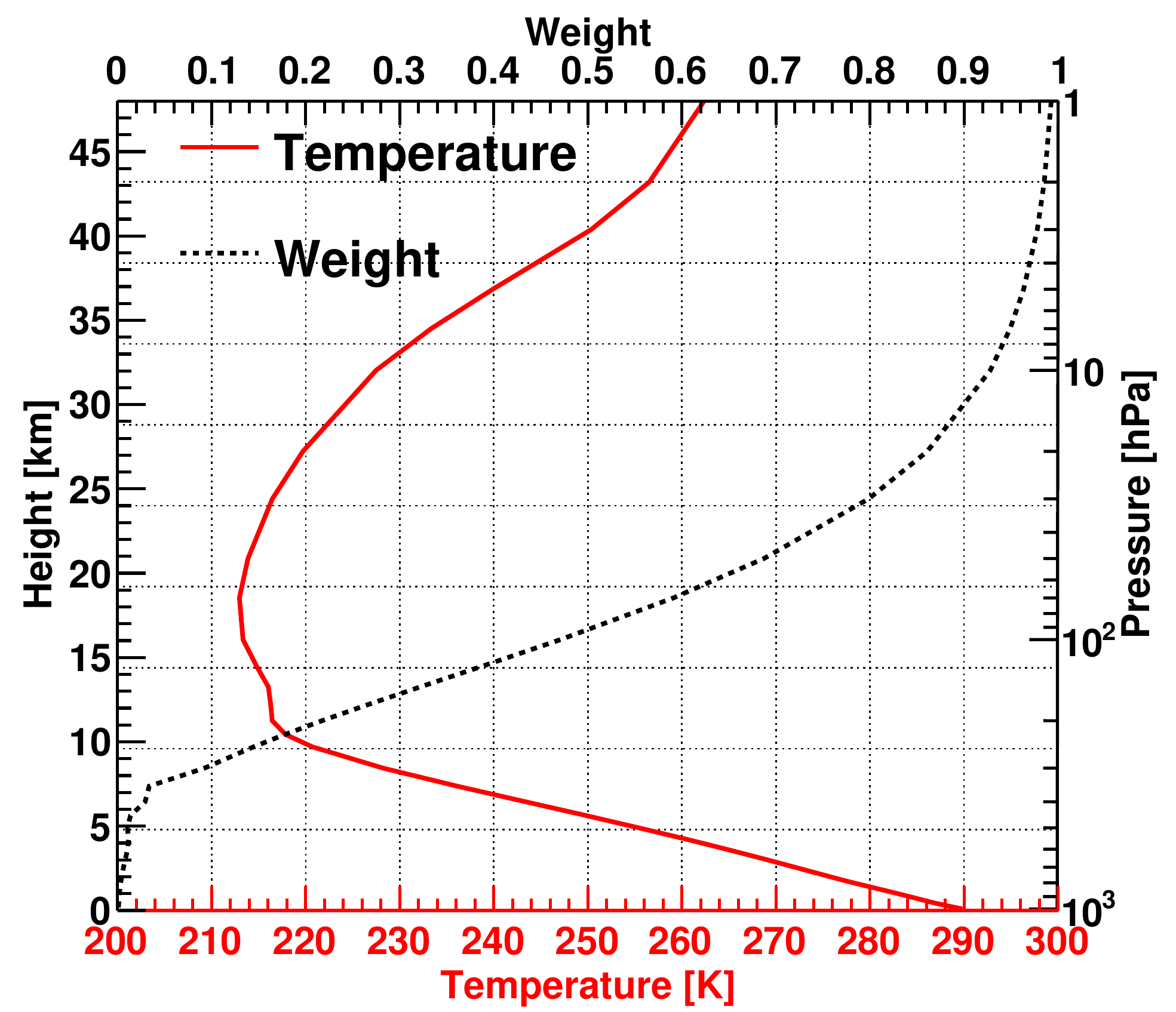}
\vspace{0.0cm}
\caption{Solid red line: mean daily temperature profile, averaged over the entire data set, as a function of pressure. The pressure range is from 1000 hPa, near Earth's surface, to 1 hPa, near the top of the stratosphere. Dashed black line: weight as a function of pressure, used to calculate $\mathrm{T_{eff}}$ at the LNGS site (see text).}
\label{fig:tempweight}       % Give a unique label
\end{figure*}

The weight function W(X) depends on the attenuation lengths of the cosmic ray primaries, pions and kaons, and their critical energies, on the muon spectral index, on the $\mathrm{K/\pi}$ ratio, on the energy required for a muon to survive to a particular underground depth, $\mathrm{E_{thr}}$, and on the zenith angle, $\mathrm{\theta}$, of the muon. 
We have calculated $W(X)$ using the definition 
%of equations 8 and 9 
in \cite{2010Minos} and with the values of the parameters as in Table I of the same work. In turn, as the value of $\mathrm{<E_{thr}\cdot cos \theta >}$ is site-dependent, we have performed its calculation for LVD. To do so, we have generated $\mathrm{1\cdot10^6}$ muons with the MUSIC and MUSUN simulation codes \cite{MUSIC},\cite{MUSUN}, which take into account the rock density and distribution around the LNGS \cite{MUSIC}, obtaining the energy and angular distribution underground.
$\mathrm{E_{thr}}$ has been calculated for each muon accounting for the rock overburden corresponding to its incoming direction. We have then checked if the muon would generate a trigger in LVD in its nominal configuration. For all the muons that satisfy the trigger condition, we have included the corresponding value of $\mathrm{E_{thr} \cdot  cos \theta}$ in the calculation of the average. The obtained value of $\mathrm{<E_{thr}\cdot cos \theta >}$ is 1.40 TeV, which is the value that we adopt in this work. This value is different from the one used by other experiments at Gran Sasso \cite{2012Bellini}\cite{2016Agostini} (1.833 TeV), as in that case it represented the energy threshold at depth of 3400 m w.e. as adopted in \cite{2010grashorn}.
The density of the rock in the Gran Sasso mountain is known with a systematic uncertainty of 2\% which results into an uncertainty of 0.05 TeV in $\mathrm{<E_{thr} \cdot cos \theta >}$. To be conservative and to account for other possible uncertainties, namely those related to the distribution of the rock, difficult to estimate with precision, we consider in the following a systematic uncertainty of 5\% on the rock density corresponding to an uncertainty of 0.13 TeV in $\mathrm{<E_{thr} \cdot  cos \theta >}$. Figure \ref{fig:tempweight} shows the weight function used in this work as a function of pressure level in the atmosphere (dashed black line), in the range 1-1000 hPa, i.e., from the Earth surface up to nearly 50 km.

\begin{figure*}[htbp]
\centering
\vspace{-0.0cm}
\includegraphics[width=8cm,clip]{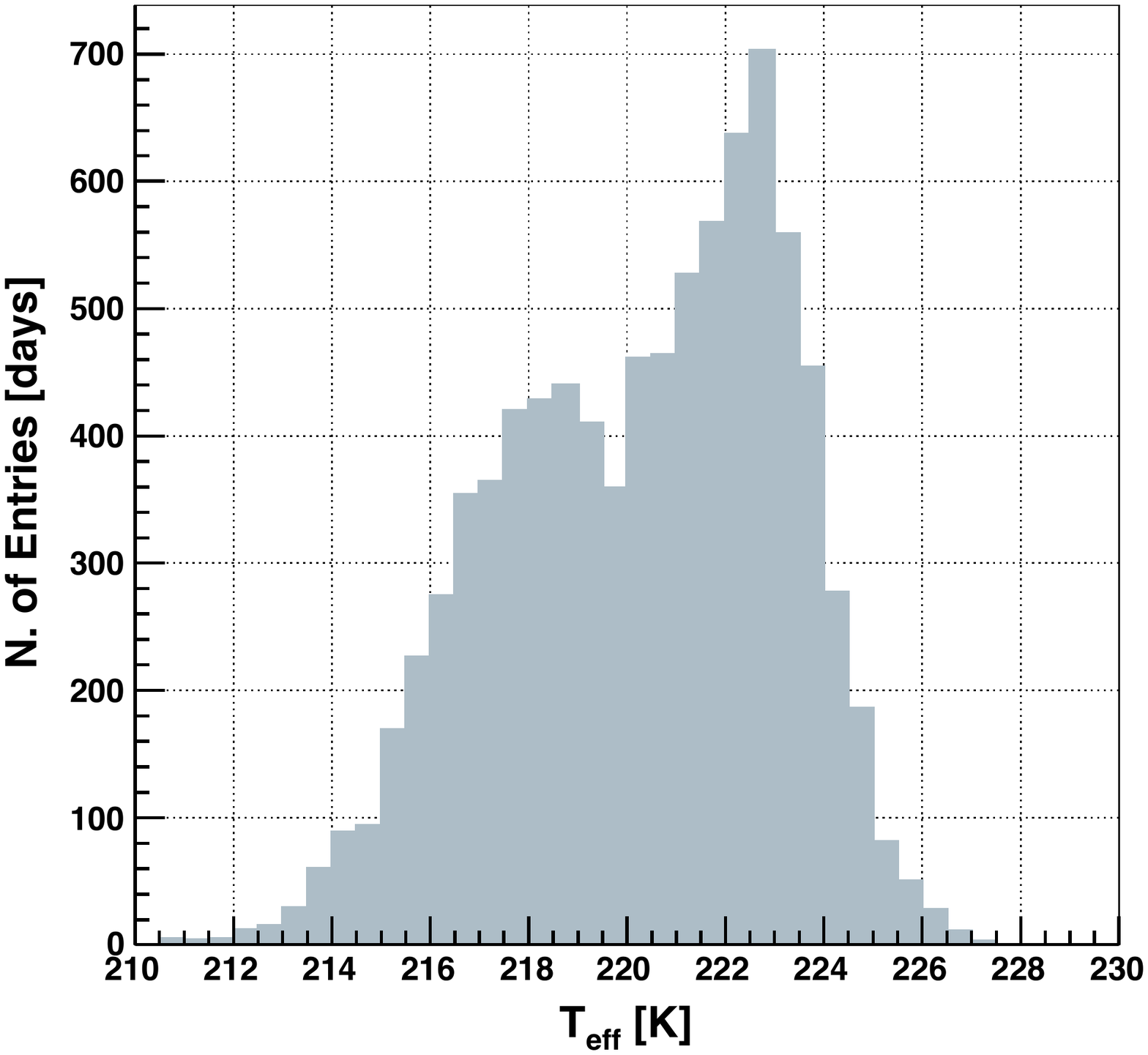}
\vspace{-1.5cm}
\caption{Distribution of the daily effective temperature over the period considered in this work. The average is $\mathrm{T^0_{eff} = 220.307\pm 0.006~K}$.}
\label{temp}       % Give a unique label
\end{figure*} 

%We note that in this work we calculate $\mathrm{T_{eff}}$ independently for the four data sets available in each day. The four values are then averaged and their variance is used to estimate the uncertainty on the mean value. The distribution of the daily effective temperature over the period considered in this work is shown in Figure \ref{temp}, the average being $\mathrm{T^0_{eff} = 220.307\pm 0.006~K}$.
We note that in this work we calculate the effective temperature independently for the four data sets available for each day. The four values are then averaged and their variance, typically 0.5 K, is used to estimate the uncertainty on the mean value. The distribution of the daily effective temperature over the period considered in this work is shown in Figure 2, the average being $\mathrm{T^0_{eff}=220.3}$~K. 
It is worth to remark that the distribution is bimodal (with peaks at 218.3 K and 222.6 K) and asymmetric with respect to the mean value. The former characteristic is caused by the presence of the annual temperature modulation, while the latter reflects the fact that such modulation is not purely sinusoidal, also due to the SSW events. These phenomena, which take place during winter in the northern hemisphere, are marked by sudden and fast increases of temperature.

To evaluate the systematic uncertainty due to the use of the ECMWF model, the temperature data were cross-checked, for the period 2002-2017, using measurements from the AIRS instrument \cite{AIRS} onboard the NASA AQUA satellite \cite{AQUA}. Launched in 2002, AIRS is an infrared sounder providing the temperature profiles in the atmosphere twice a day at the selected location. The differences between the daily ECMWF and AIRS effective temperatures are well described by a Gaussian distribution with $\sigma=0.7$~K. We consider the latter, added in quadrature to the daily variance in the ECMWF model, as the total systematic uncertainty on the effective temperatures, corresponding to 0.9~K.

\section{The muon data set}
\label{sec.mu}

LVD is a 1000 t liquid scintillator instrument aimed at detecting neutrinos from core collapse supernovae \cite{2015Agafonova}.
Given its goal, one of its essential features is its modularity: it consists of an array of 840 scintillator counters, organised in sub-sectors that can take data independently one from another. Such modular structure allows LVD to achieve a duty cycle close to 100\%. 

%Added ex-referee the fact that LVD was completed in 2001.
Another crucial feature is its long-term operation: LVD has been continuously taking data for a quarter century, since June 1992, its mass increasing from 300 t to the final one of 1000 t in January 2001. The two features make the instrument a very appropriate one to continuously study the underground muon flux and investigate its variations.
In this work, we use data from January 1994 to December 2017: over this period LVD was active for 8659 days, corresponding to 99\% livetime (see Table \ref{t1}). 
Data collected in the first one year and half (1992 - 1993) are not used in the following analysis because of frequent interruptions in the data taking in the early phases of operation. 

A detailed description of the instrument is given in \citep{2015Agafonova}: we recall here the main characteristics related to the selection of muons in the scintillator detector\footnote{
Between 1992 and 2002, LVD was equipped also with muon-tracking detectors, namely limited
streamer tubes, which surrounded two faces of the scintillator counters. Data from those detectors were
exploited for different muon studies, such as those in \cite{aglietta1994}, \cite{1998Aglietta} and \cite{1999Aglietta}, but are not used in this work to ensure a uniform approach over the whole 24-years data set.}.
Each 1.5 m$^{3}$ scintillator counter is viewed from the top by three photomultipliers (PMTs). The LVD trigger logic (extensively described in \cite{2007Agafonova}) is based on the 3-fold coincidence of the PMTs in a single counter and corresponds to an electron-like energy-release threshold well below 10 MeV. The energy resolution of the counter, at 10 MeV, is $\mathrm{\sigma / E \sim 20 \%}$. The time of occurrence of each event is measured with a relative accuracy of 12.5 ns and an absolute one of 100 ns.

In this work, muons are identified through the time coincidence of signals with energy $>10$ MeV, within 175 ns, in two or more counters (this time width is chosen to take into account for the jitter of the PMT's transit time). We apply to individual counters the same quality cuts that have been described in \citep{2015Agafonova}, based on checks of their counting rate and energy spectrum. The average rate of muons crossing LVD is monitored and it is $\mathrm{0.097 \pm 0.010~s^{-1}}$, the mean per counter being $\mathrm{f_\mu(c)~\sim50~d^{-1}}$. Account taken for the number of counters as well as of days of operation, we consider that a cut at 5 s.d. in the rate is adequate to reject the malfunctioning ones: we reject those whose rate is smaller than $15~\mathrm{d}^{-1}$ or larger than $85~\mathrm{d}^{-1}$. An anomalous muon rate is primarily due to hardware problems, either in the scintillator, or in the PMTs or in the electronics. The percentage of counters rejected by this cut is about 5\%.  We check also the energy spectrum in each counter, i.e., the distribution of energy losses of muons. While the above described rate-based cut rejects rather naturally also all counters which show an anomalous spectrum, the aim of a further check on the spectrum is to verify the counter calibration. Given the low daily rate of muons, the energy spectrum is built every month for each detector.  This is compared, through a $\chi^2$ test, with a reference one, obtained through a full Monte Carlo simulation. The number of rejected counters due to this selection alone is usually few over the total 800. 
The sequence of quality cuts ends with a check of the daily counting rate above a lower threshold, namely 7 MeV: the larger statistics allows us to identify, at the trigger level, noisy or unstable counters. We require that the daily counters counting rate at $\mathrm{E}\ge 7$~MeV is lower than $\mathrm{3 \cdot 10^{-3} s^{-1}}$. This cut affects on average 2\% of the counters.

Finally, we discard muon events produced by the CNGS neutrino beam (CERN Neutrino to Gran Sasso \cite{CERN2}), which was active between 2006 and 2012. 
Namely, all events occurring in a veto window ($\mathrm{\pm~20~\mu s}$) set around the CNGS spill (duration $\mathrm{10.5~\mu s}$) are discarded. The additional dead time, due to  the $10^7$ spills in the period 2006-2012, is $\mathrm{\sim 500~s}$, corresponding to $\sim 50$~ cosmic muons lost from the analysis \cite{LVDNIM} \cite{nuspeed}. 

\begin{table}[b]%The best place to locate the table environment is directly after its first reference in text
\caption{\label{t1}%
Quality cuts applied to the
events. $\epsilon$ stands for the
overall efficiency. The explanation for the different cuts can be found
in the text.\\}

\begin{ruledtabular}
\begin{tabular}{lcdr}
%\multicolumn{4}{c}{LVD The muon data set } \\
%\hline
\multicolumn{1}{l}{} & \multicolumn{1}{c}{days} & \multicolumn{1}{c}{$\mathrm{N_{\mu}/10^{6}}$} &\multicolumn{1}{c}{$\mathrm{\epsilon}$(\%)}
\tabularnewline
\hline\\
\multicolumn{1}{l}{Time (from Jan-1-1994 to Dec-31-2017)} & \multicolumn{1}{c}{8766} & &\multicolumn{1}{c}{100} 
\tabularnewline
\multicolumn{1}{l}{Live Time} & \multicolumn{1}{c}{8659} &  \multicolumn{1}{c}{55.8}&\multicolumn{1}{c}{98.8} 
\tabularnewline
\multicolumn{1}{l}{After quality cuts rejection } & \multicolumn{1}{c}{8543} &  \multicolumn{1}{c}{55.4}& \multicolumn{1}{c}{97.5} 
\tabularnewline
\multicolumn{1}{l}{$\mathrm{I_{\mu}^0 \times (0.925) \le I_{\mu} \le I_{\mu}^0 \times (1.075)}$} & \multicolumn{1}{c}{8402} & \multicolumn{1}{c}{54.8} &\multicolumn{1}{c}{95.8} 
\tabularnewline
\end{tabular}
\end{ruledtabular}
\end{table}

\begin{figure*}[htbp]
\centering
\vspace{-1.5cm}
\includegraphics[width=14cm,clip]{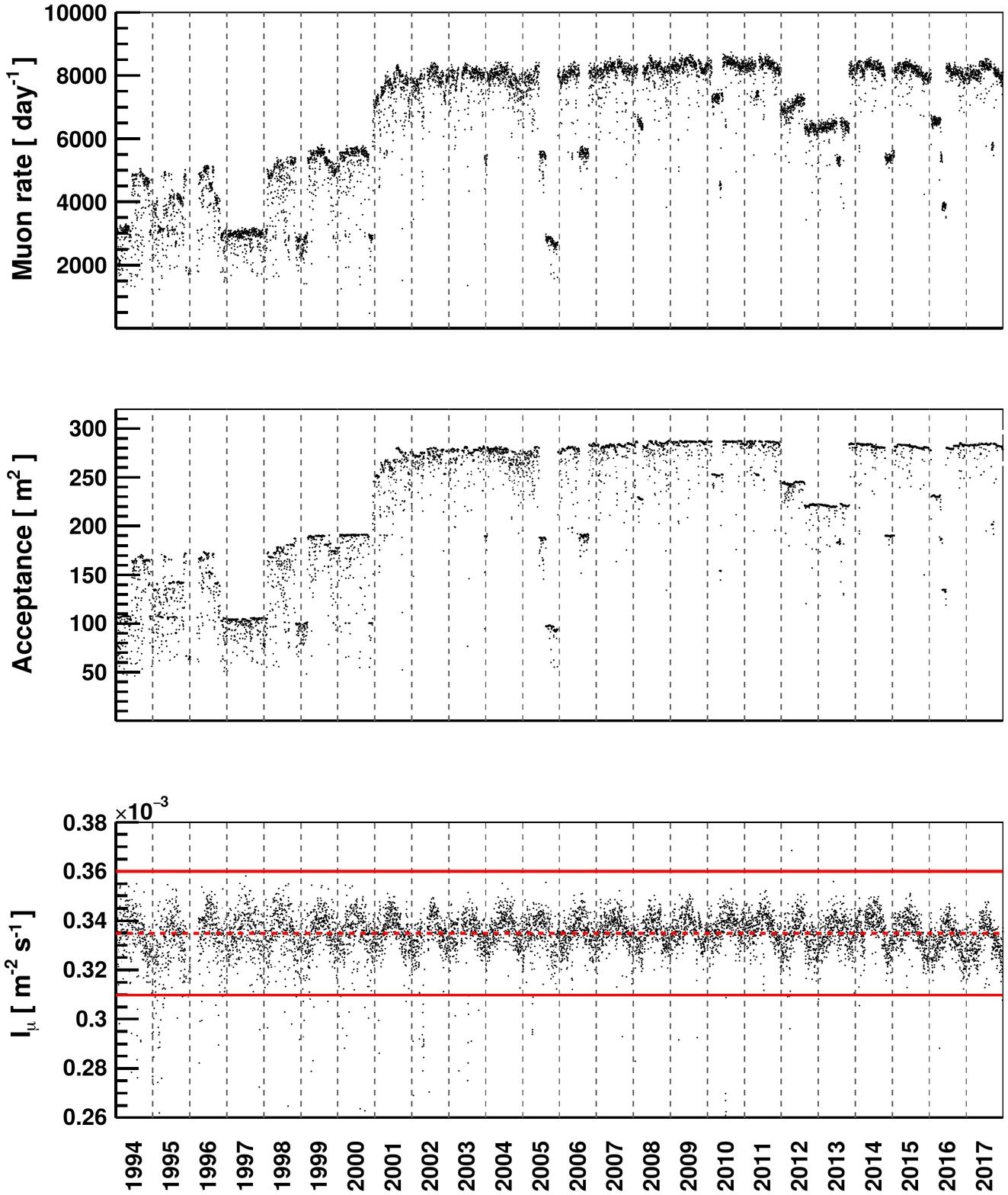}
\vspace{0.0cm}
\caption{Top panel: muon rate ($\mathrm{day^{-1}}$), after the quality cuts (see text) as a function of time, from 1994 to 2017. Middle panel: detector acceptance as a function of time. Bottom panel: muon flux as a function of time. The dashed red line corresponds to the average flux, while the two red solid lines represent the flux limits within which data are used (see text).
}
\label{tripla}       % Give a unique label
\end{figure*} 

After applying the quality cuts and subtracting the CNGS muons, the data set consists of $\mathrm{5.54 \times 10^7}$ muons for a total of 8543 live days as shown in Table \ref{t1}.
The number of muons per day is shown as a function of time in the top panel of Figure \ref{tripla}.
The observed behaviour of the rate is due to the varying acceptance of the detector over time. As LVD is a modular detector its configuration can vary over time, due to, e.g., deployment, or maintenance, or temporary problems of part of the scintillator counters. 
The list of active and well-functioning counters is determined day-by-day. To properly take into account in the calculation of the acceptance all the configurations and their time variability, we have developed a detailed Monte Carlo simulation of the detector with the GEANT4 toolkit \cite{2003GEANT}.
The distribution of the muon energies and arrival directions is generated accordingly to the MUSIC and MUSUN codes \cite{MUSIC},\cite{MUSUN}, developed for the Gran Sasso rock distribution around the LNGS. 
For each selected direction, muons are generated uniformly over a large circle centered in the middle of LVD, with radius large enough to contain the whole detector. 
Muons are then tracked through LVD: the information on the number of crossed counters, together with the arrival time and the energy released in each counter, are stored. 
To define a muon event, we apply to the output of the Monte Carlo simulation the same muon-selection cuts previously described. First, we generate 100000 muons through the detector in its nominal configuration 
i.e., with all scintillators counters active. We take the corresponding acceptance, averaged over the cosmic muon arrival directions in the LNGS, as a reference: it results to be $\mathrm{(298 \pm 3)~m^2}$.
%Removed ex-referee: $\mathrm{S_{o} = (298 \pm 3)~m^2}$.
We then throw the muons on the detector simulating on a daily basis each real configuration, as obtained after applying the quality cuts on the counters.
%including the quality cuts. 
We finally calculate the daily relative acceptance as the ratio between the number of muons detected with each configuration and that detected with the reference one.
We show, in Figure \ref{tripla}, middle panel, the resulting daily acceptance as a function of time in the considered data period.
The associated uncertainty is about 1\%: it is mostly due to the systematics associated to the muon direction given by the MUSUN code.\\
The calculation of the exposure allows us to derive the muon flux as the ratio between the number of muons and the acceptance.
%Addition ex-referee on 1994-2000.
The muon flux is shown as a function of time in Figure \ref{tripla}, bottom panel. The dashed red line represents the average ($\mathrm{3.34 \pm 0.0005^{stat}\pm 0.03^{sys} \cdot 10^{-4} ~m^{-2} s^{-1}}$). The larger fluctuations in the period from 1994 to 2000 are due to the fact that the array was taking data with a lower active mass, hence with a smaller acceptance.

%Removed ex-referee and modified. One can notice that, especially in that period, there are points corresponding to fluxes significantly lower than the average, as can also be seen in the distribution shown in Figure \ref{fig:daily}. These are due to an overestimation of the acceptance caused by a misclassification of the counters status. Therefore, we exclude from the analysis the days where the muon flux variations with respect to the average are greater than 7.5$\%$: the two solid red lines in the figure represent the flux limits within which data are used in the following. After this cut, the number of days in the data set is reduced by $\mathrm{1.7 \%}$. It is interesting to note that the muon flux distribution is not bimodal, differently from the temperature one (see Fig. \ref{temp}); this can be ascribed to the effect of the higher fluctuations (statistical and systematic) present in the muon flux series.\\

\begin{figure*}[htbp]
\centering
\vspace{-0.0cm}
\includegraphics[width=9cm,clip]{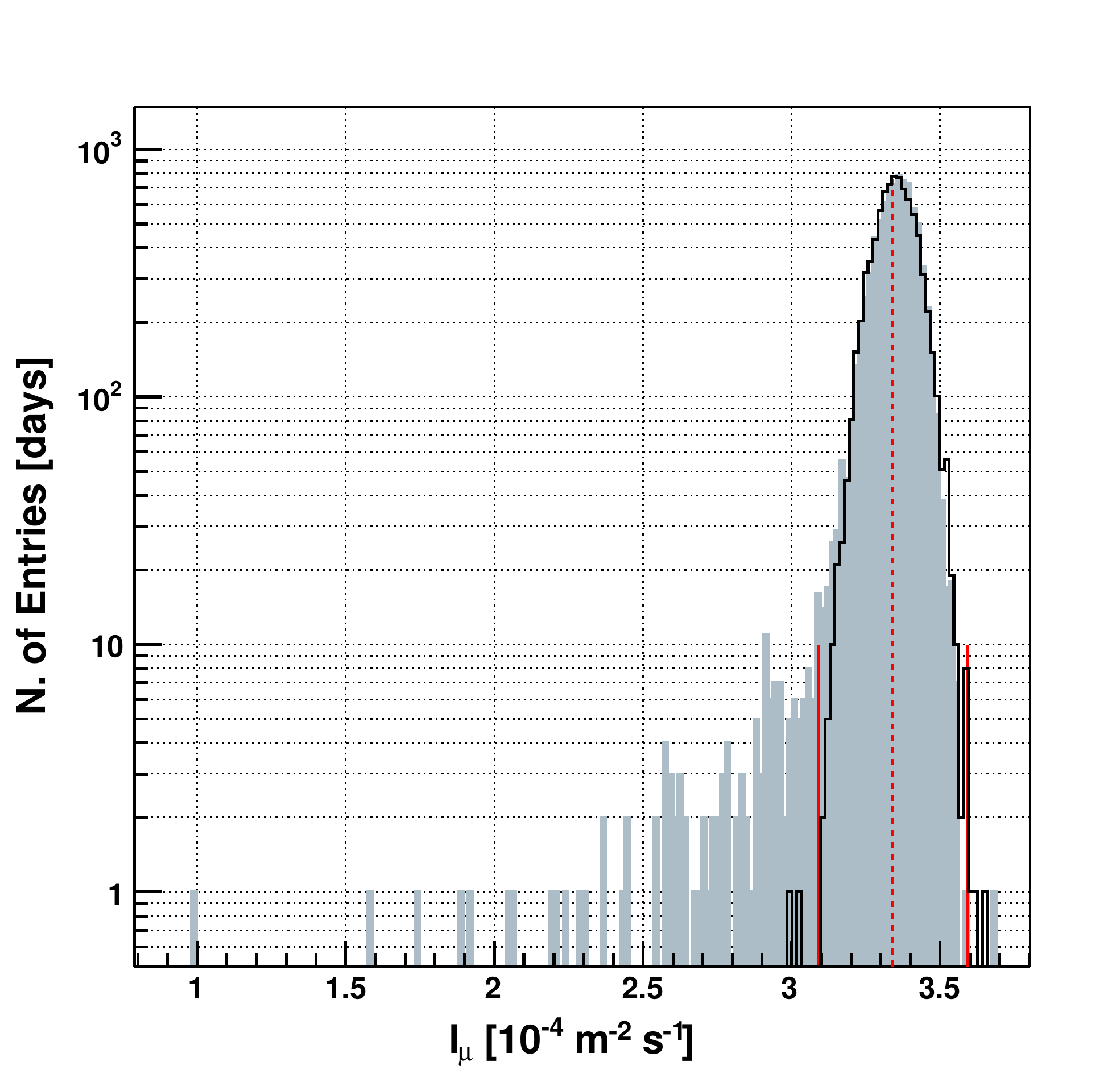}
\vspace{0.0cm}
\caption{Distribution of the daily muon flux (grey histogram). The dashed red line represents the mean value. The black histogram represents the distribution expected from the temperature variations taking into account the expected correlation coefficient as well as the statistical and systematic uncertainties on the muon flux. The two continuous lines define the flux limits within which data are used.}
\label{fig:daily}       % Give a unique label
\end{figure*}

One can notice that, especially in the first half of the data set when the detector was under construction and commissioning, there are points corresponding to fluxes significantly lower than the average, as can also be seen in the distribution shown in Figure \ref{fig:daily}. These outliers are due to instrumental effects, namely to an overestimation of the acceptance caused by a misclassification of the counters status. To define sensible cuts, we build the expected distribution of the muon flux by folding the temperature distribution scaled with the expected correlation coefficient at LVD depth, 0.90, with the statistical and systematic uncertainties of the flux, summed in quadrature. 
The resulting distribution (shown as a black line in Figure \ref{fig:daily}) is well-fit by a Gaussian curve whose width is 2.5\%. The fact that it is not bimodal, differently from the temperature one (see Fig. \ref{temp}) is due to the effect of the higher fluctuations (statistical and systematic) present in the muon flux series. We then exclude from the analysis the days when the muon flux variations with respect to the average are greater than 7.5\%, i.e., 3 s.d.: the two solid red lines in the figure represent the flux limits within which data are used in the following. After this cut, the number of days in the data set is reduced by 1.7\%.

%
% (see Fig. \ref{fig:ssa_0}).
In conclusion, the data set used in the following consists of $\mathrm{5.48 \times 10^7}$ muons collected over 8402 days. The sequence of applied cuts is shown in Table \ref{t1}. The obtained average muon flux is $\mathrm{I_{\mu}^0 = 3.35 \pm 0.0005^{stat}\pm 0.03^{sys} \cdot 10^{-4} ~m^{-2} s^{-1}}$, consistent with previously obtained measurements by other detectors in the same laboratory \cite{1997MACRO},\cite{2012Bellini},\cite{{2016Agostini}}\cite{2017Opera}\cite{2019Agostini}.

\section{Correlation between the muon flux and the temperature}
\label{corr}
\begin{figure*}[!htbp]
\centering
%\vspace{2.5cm}
%\hspace{-1.5cm}
\includegraphics[width=14cm,clip]{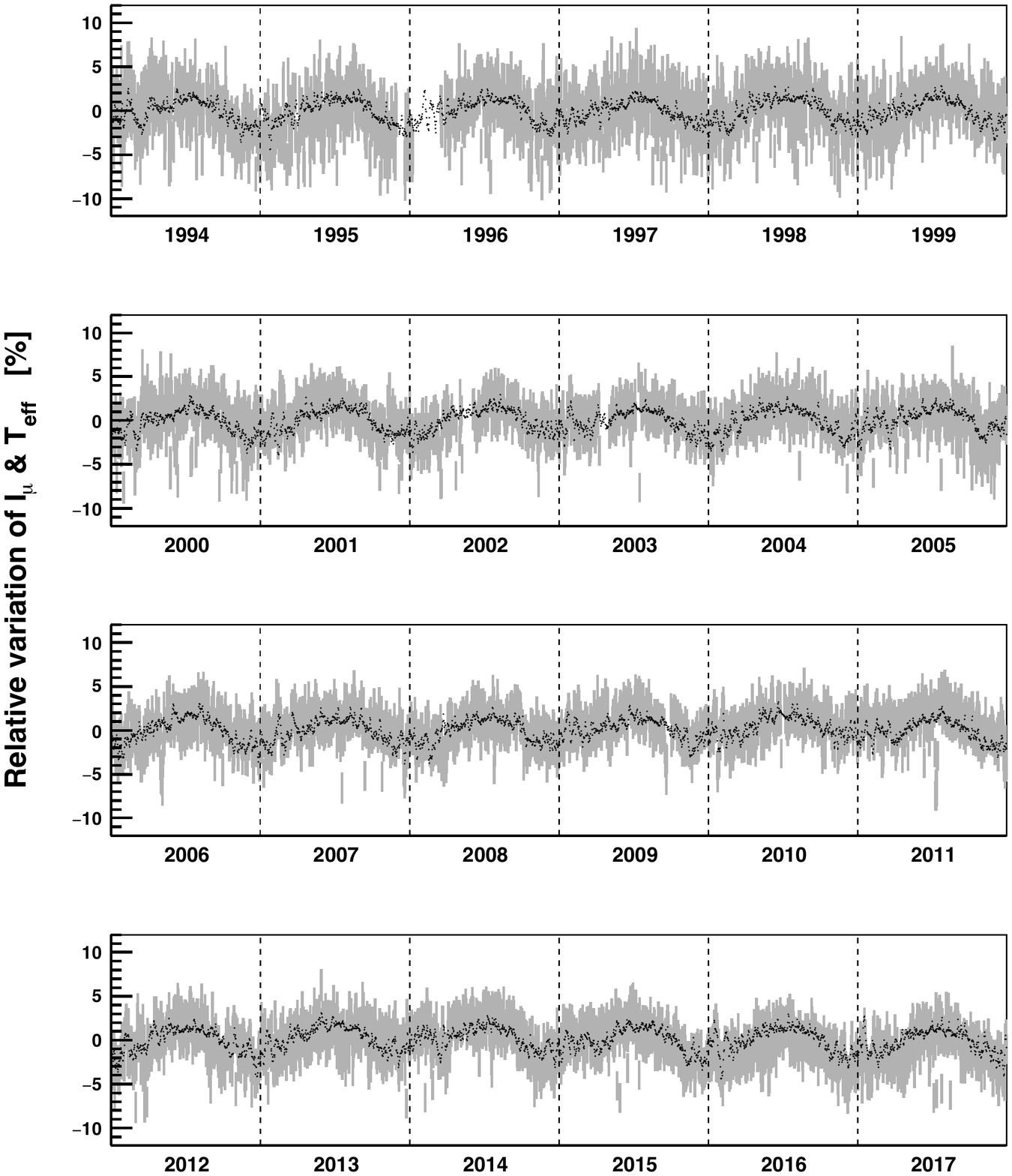}
\vspace{-2.0cm}
\caption{Grey histogram: percent variations of the daily muon flux, $\mathrm{\Delta I_{\mu}/I_{\mu}^0}$ as a function of time. The error bars represent the statistical uncertainty. Black points: percent variations of the daily effective temperature, $\mathrm{\Delta T_{eff}/T_{eff}^0}$, as a function of time.}
\label{fig:2}       % Give a unique label
\end{figure*}

\begin{figure*}[htpb]
\centering
%\vspace{-2.0cm}
\includegraphics[width=10.cm,clip]{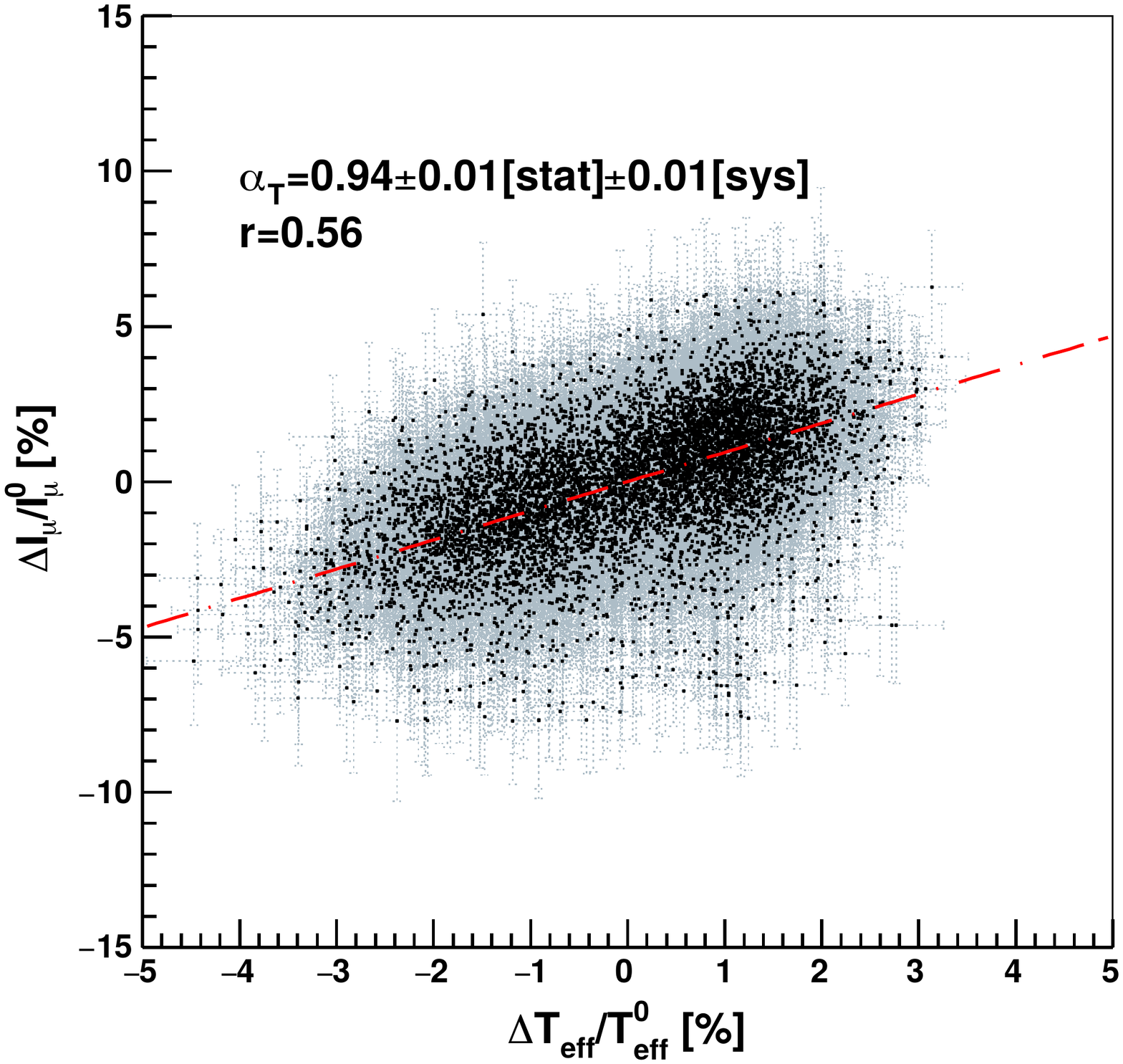}
%\vspace{-0.5cm}
\caption{Correlation between muon-flux and temperature variations, $\mathrm{\frac{\Delta I_{\mu}}{I_{\mu}^0}~ vs ~\frac{\Delta T_{eff}}{T_{eff}^0}}$, together with the resulting linear fit (red dashed line).} 
%{\bf{with the $\mathrm{3\sigma}$ band.}}}
\label{fig-000}       % Give a unique label
\end{figure*}

We study in this section the correlation of the flux of the muons selected as described in Section \ref{sec.mu}, with the effective temperature, derived as detailed in Section \ref{sec.T}. \\
As explained in the introduction, an increase in the atmospheric temperature should lead to an increase in the observed muon rate: a positive correlation is hence expected and it is observed in our data, as can be seen in Figure \ref{fig:2}.
The grey histogram and the black points show, respectively, the relative deviations from the mean daily muon flux, $\mathrm{\Delta I_{\mu}/I_{\mu}^0}$, and from the mean temperature, $\mathrm{\Delta T_{eff}/T_{eff}^0}$, as a function of time. The correlation between the two data sets is evident. We calculate the effective temperature coefficient, $\mathrm{\alpha_{T}}$, as:
\begin{equation}
\mathrm{\frac{\Delta I_{\mu}}{I_{\mu}^0}} = \mathrm{\alpha_{T}} \mathrm{\frac{\Delta T_{eff}}{T_{eff}^0}}
\end{equation}
A linear regression provides us with the value of $\mathrm{\alpha_{T}}$ which results to be $0.94 \pm 0.01 \mathrm{(stat)} \pm 0.01 \mathrm{(syst)}$, with the strength of the correlation being $0.56$ for 8402 data points. The actual correlation between muon-flux and temperature variations is shown in Figure \ref{fig-000}, black points, together with the resulting linear fit (red dashed line). 

%Inserted ex-referee: systematic uncertainties
\begin{table}[b]%The best place to locate the table environment is directly after its first reference in text
\caption{\label{xxx}%
Systematic errors on the parameter inputs to $\mathrm{\alpha_{T}}$\\}

\begin{ruledtabular}
\begin{tabular}{lcdr}
%\multicolumn{4}{c}{LVD The muon data set } \\
%\hline
\multicolumn{1}{l}{} & \multicolumn{1}{c}{value} & \multicolumn{1}{c}{$\mathrm{\Delta \alpha_{T}}$} %&\multicolumn{1}{c}{$\mathrm{\epsilon}$(\%)}
\tabularnewline
\hline
\multicolumn{1}{l}{Meson production ratio, $\mathrm{r_{K/\pi}}$} & \multicolumn{1}{c}{$\mathrm{0.149\pm0.06}$ from \cite{2010Minos}} & \multicolumn{1}{c}{0.002} 
\tabularnewline
\multicolumn{1}{l}{Mean effective temperature} & \multicolumn{1}{c}{$\mathrm{220.3\pm0.9~K}$ our calculation} & \multicolumn{1}{c}{0.004} 
\tabularnewline
\multicolumn{1}{l}{Threshold energy, $\mathrm{{<E_{thr} \cdot cos\theta>}}$} & \multicolumn{1}{c}{$\mathrm{1.40\pm0.13~TeV}$ our calculation} & \multicolumn{1}{c}{0.002} 
\tabularnewline
\multicolumn{1}{l}{LVD acceptance} & \multicolumn{1}{c}{$\mathrm{298\pm3~m^2}$ our simulation} & \multicolumn{1}{c}{0.01} 
\tabularnewline
\hline
\multicolumn{1}{l}{\bf{Total systematic error budget}} & & \multicolumn{1}{c}{\bf{0.011}} 

\tabularnewline
\end{tabular}
\end{ruledtabular}
\end{table}

\begin{figure*}[htbp]
\centering
%\vspace{0.cm}
\includegraphics[width=10cm,clip]{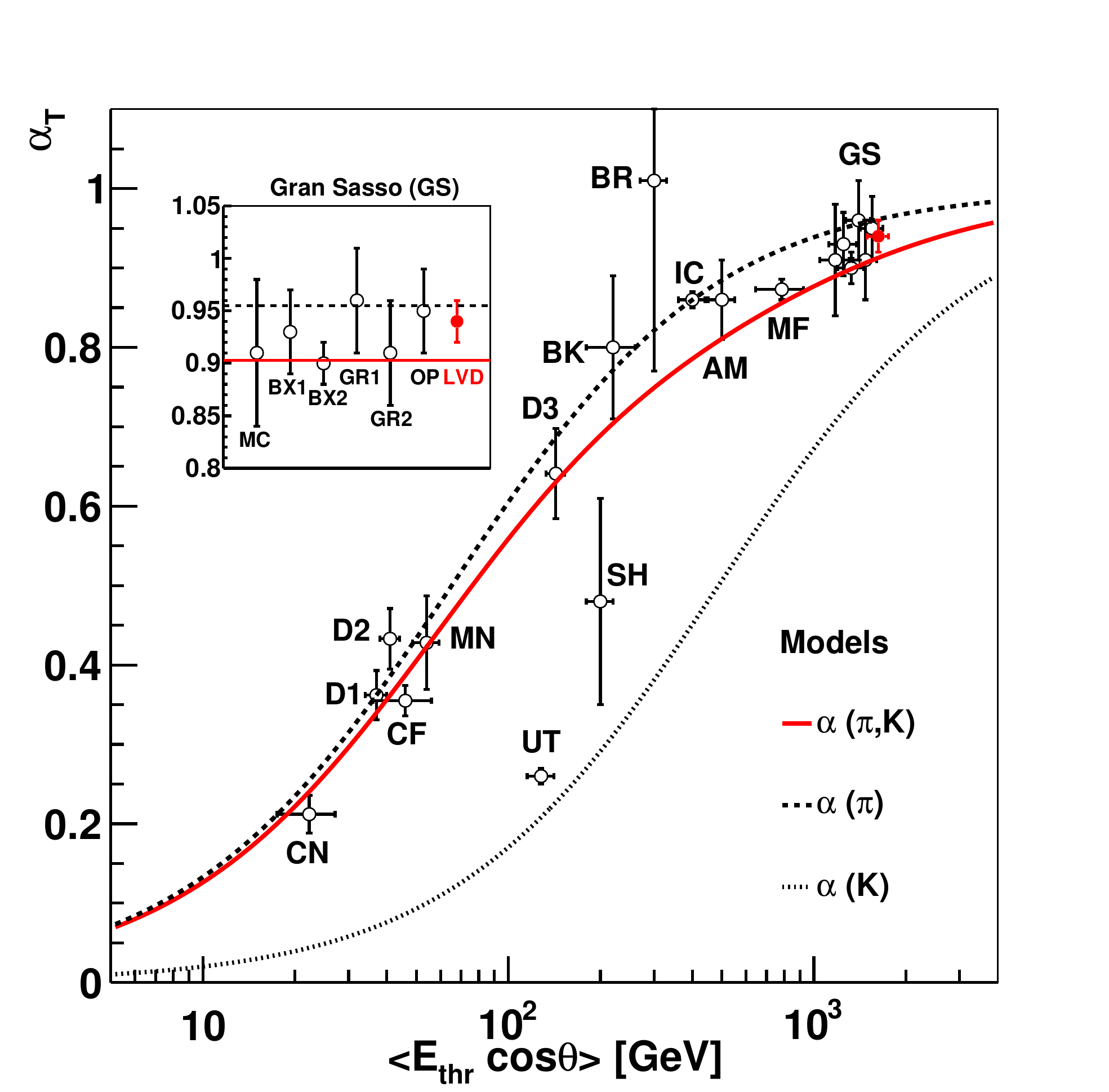}
\caption{Comparison of the experimental $\mathrm{\alpha_{T}}$ values with the models~\cite{2010grashorn} accounting for pions and kaons decays (solid red line), pions decays only (dashed black line) and kaons decays only (dotted black line). The value determined in this work is reported as a filled red point, the error bar corresponding to the sum of the statistical and systematic uncertainties. The open points represent the values determined by other experiments: Amanda (AM) \cite{1999Amanda}, Baksan (BK) \cite{1991Baksan}, Barrett (BR) \cite{1954Barrett}, the three experimental halls of Daya Bay (D1, D2 and D3) \cite{2018DayaBay}, Icecube (IC) \cite{2011Icecube}, MINOS Near (MN) \cite{2014Minos} and Far (MF) \cite{2010Minos} detectors, Double Chooz Near (CN) and Far (CF) detectors \cite{2017DoubleChooz}, Sherman (SH) \cite{1954Sherman}, and Utah (UT) \cite{1981Utah}. The six Gran Sasso (GS) based measurements are highlighted in the inset and include MACRO (MC) \cite{2003Macro}, Borexino (BX1 and BX2) \cite{2012Bellini},\cite{2019Agostini}, GERDA (GR1 and GR2) \cite{2016Agostini}, Opera (OP) \cite{2017Opera} and LVD (this work). 
They are artificially displaced on the horizontal axis for a better visualization.}
\vspace{0.cm}
\label{fig:t}       % Give a unique label
\end{figure*}

The sources of the systematic uncertainty associated to the measurement of $\mathrm{\alpha_{T}}$, summarised in Table \ref{xxx}, are, on the one hand, the LVD acceptance, which enters into the calculation of the muon intensity, and, on the other hand, the weight function $W(X)$, which enters into the calculation of the effective temperature. The former, which has a systematic uncertainty of 1\% (see Section \ref{sec.mu}), gives the largest contribution to the total budget. The systematic uncertainty on the latter has in turn three main sources: the meson production rate, the calculation of  $\mathrm{<E_{thr} \cdot cos\theta>}$ and that of the mean effective temperature. Note that the uncertainty on the $\mathrm{K/\pi}$ decay constants are also a source of uncertainty but, given that in \cite{2010Minos} they have been shown to have a subdominant effect, they are not included in the table. The values of these systematic uncertainties shown in the table are evaluated by modifying each of the parameters used in the analysis by their uncertainty and by recalculating again $\mathrm{\alpha_{T}}$ value. Table \ref{xxx} shows the deviations found with respect to the central value of $\mathrm{\alpha_{T}}$: one can note that the most important source of systematics for the weight function is the calculation of the mean effective temperature. The total systematic uncertainty, obtained by adding in quadrature all the contributions, amounts to 0.01, reflecting that on the acceptance. The estimated uncertainty has been validated by performing a cross-check on the stability of the measurement, namely by using only data taken during the period when LVD was complete, between 2001 and 2017. The obtained value of $\mathrm{\alpha_{T}}$ is consistent within 1 s.d. with that found using the full data set.\\
Figure \ref{fig:t} shows how the coefficient  $\mathrm{\alpha_{T}}$ measured in this work (filled red point) compares with those measured by other underground experiments (open points) and with model predictions (lines)~\cite{2010grashorn}. The solid red line represents the prediction including the contributions of pion and kaon decays, while the dashed and dotted lines account for one single production mechanism only, pions and kaons decay, respectively. All the experimental values are presented as a function of $\mathrm{<E_{thr} \cdot cos\theta>}$, which is the only site-dependent parameter affecting the weight function W(X) calculation (see Section II). For experiments not quoting the corresponding $\mathrm{<E_{thr} \cdot cos\theta>}$, we determine the value and its uncertainty following the prescriptions in \cite{2018DayaBay}. 
The inset in figure \ref{fig:t} compares the $\mathrm{\alpha_{T}}$ values measured by different experiments located at the LNGS. 
One can note in particular the good agreement between the LVD measurement and those by the other experiments in the same location \cite{2003Macro}, \cite{2012Bellini}, \cite{2016Agostini}, \cite{2017Opera}, \cite{2019Agostini}, and the decrease in the uncertainty of the LVD measurement, due to the large exposure of muon data considered in this work.

\section{Spectral analysis of the muon and temperature series}
\label{spect}
In this section we aim at characterizing on a year-by-year basis the modulation of the muon flux clearly visible in Figure \ref{fig:2} (gray histogram). As one can see from the same figure, the seasonal variations of the effective temperature (black dots), which drive those of the muons, are such that maxima and minima happen at slightly different times, as expected, depending on the weather evolution year by year. Other secondary and fainter variations can in fact modulate the annual cycle, such as the SSW events, which are short-term and sudden increases happening during winter time in the northern hemisphere \cite{1952Schrhag}. Consequently, we subject the two time series to a spectral analysis to estimate the power of different frequency components.

As a first step, we determine the autoregressive models for the random noise in the two series. The partial autocorrelation function (PACF), which allows one to investigate the possible presence of internal correlations in a time series, is the most effective for identifying the order of an autoregressive model. We apply this method to the two series.

\begin{figure*}[!htbp]
\centering
\vspace{-5.cm}
\includegraphics[width=15. cm,clip]{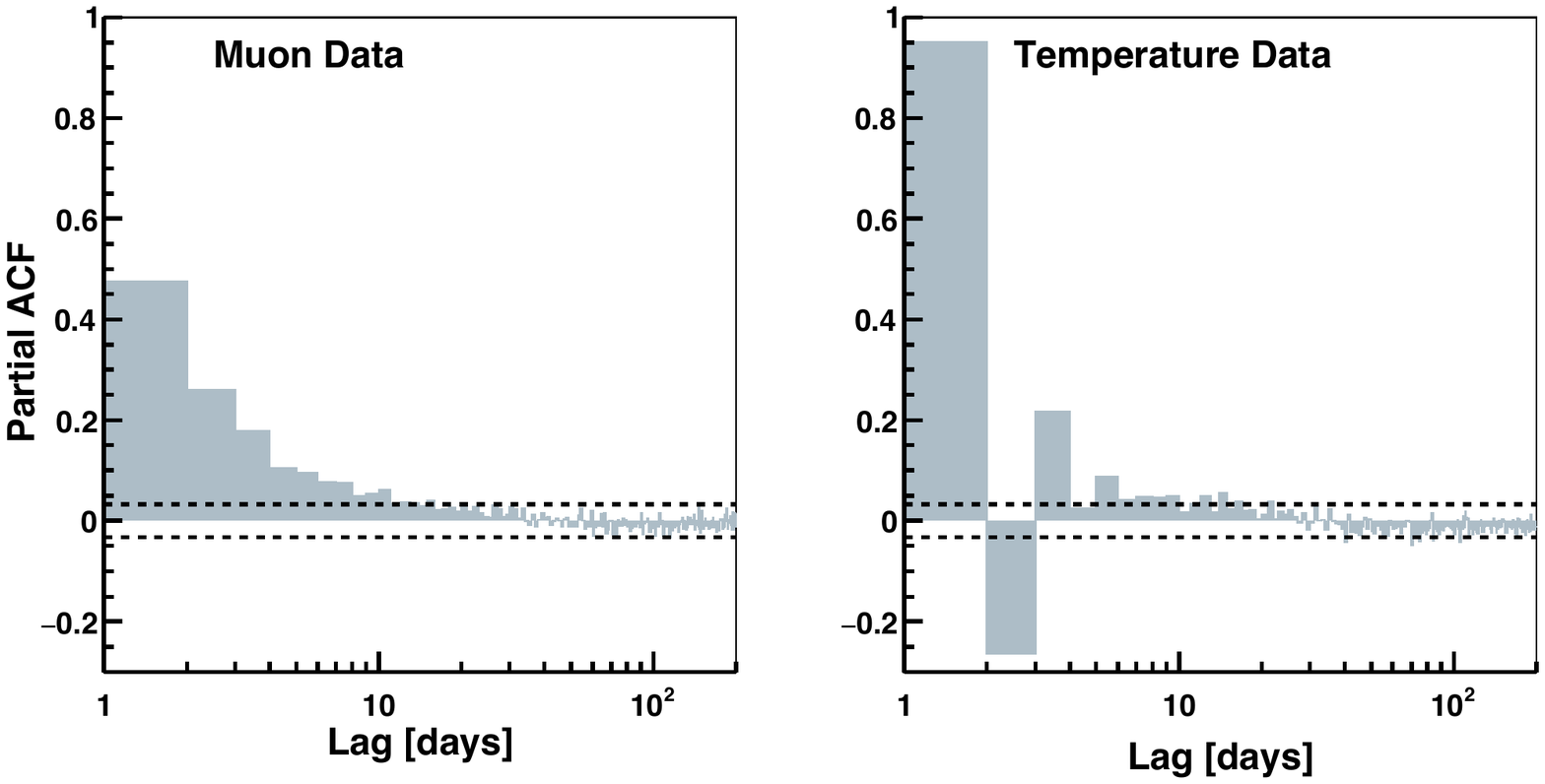}
\vspace{-5.cm}
\caption{Partial autocorrelation coefficient as a function of the time lag in days for the muon (left panel) and the effective temperature (right panel) daily time series. The dashed lines delimit the band corresponding to the 99.7\% dispersion expected from the fluctuations of a purely white noise.}
\label{fig:tt}       % Give a unique label
\end{figure*}

Figure \ref{fig:tt}, left panel, shows the partial PACF of the time-lag (in days) for the muon-flux series over a large range of time-lags. 
The dashed lines delimit the band corresponding to the 99.7\% dispersion expected from the fluctuations of a purely white noise. We find that there is an autocorrelation clearly significant above $\mathrm{3 ~\sigma}$ for lags up to about 10 days, plausibly due to the fact that the muon calibration is performed on time intervals of the same order (see section \ref{sec.mu}).
Similarly, the PACF for the temperature series is shown in Figure \ref{fig:tt}, right panel: also in this case we find that the series is significantly autocorrelated for lags up to 10 days\footnote{The possible impact of the autocorrelation on the determination of the coefficient $\mathrm{\alpha_{T}}$ has been evaluated by downsampling the two series by a factor of 10, i.e., by keeping only 1 point every 10. $\mathrm{\alpha_{T}}$ is well compatible within the statistical uncertainties.}.
These timescales of order of 10 days are longer than those of the typical baroclinic instabilities and are associated with the annular modes, which are the leading patterns of variability in the extra-tropics \cite{Thompson2000}. 

The results of the autocorrelation analysis, which exclude a pure white noise model for both time series, allow us to conclude that the random noise can be modelled in both cases by an autoregressive model of order 10 (AR10).
%Removed ex-referee: Adopting such models to describe the background, we finally investigate the spectral content of the two time series by means of the Lomb-Scargle (LS) method \cite{lomb1976least}\cite{scargle1982studies}\cite{mac1989spectral}. 
Adopting such models to describe the background, we finally investigate the spectral content of the two time series by means of the Lomb-Scargle (LS) periodogram \cite{lomb1976least}\cite{scargle1982studies}\cite{mac1989spectral}. This is a method that allows for the derivation of a Fourier-like power spectrum from a set of unevenly-sampled data, which is the case for the LVD muon-flux series.
\begin{figure*}[htbp]
%\centering
%\vspace{-2.cm}
\includegraphics[width=15. cm,clip, angle=0]{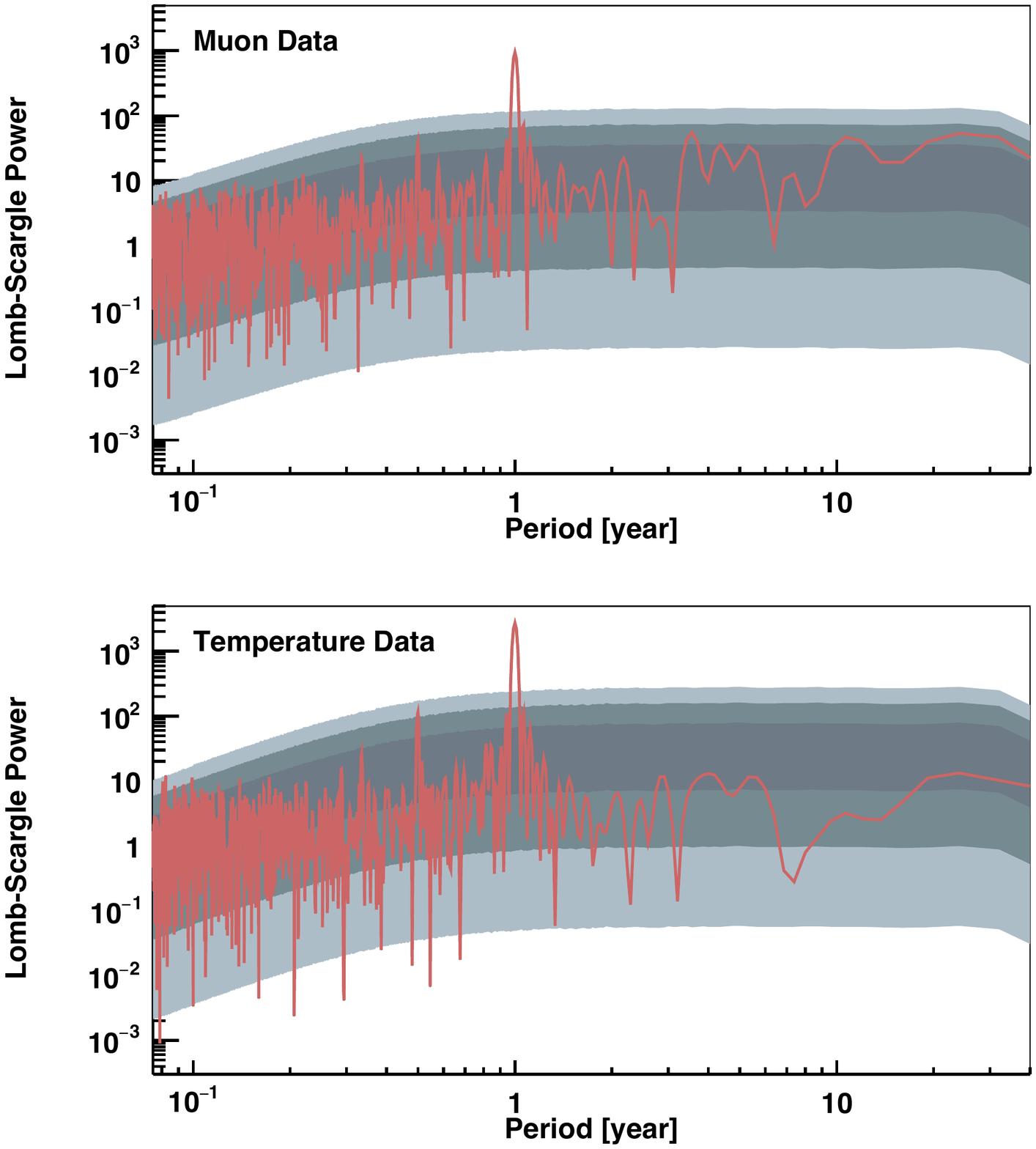}
%\vspace{-5.cm}
\caption{{Lomb-Scargle periodogram for the time series of the muon flux (top panel) and of the effective temperature (bottom panel). The three bands represent the power spectrum fluctuations at 1, 2 and $\mathrm{3 ~\sigma}$ of 10000 simulated background time series modelled by an autoregressive process of order 10.}
}
%\vspace{-2.cm}
\label{LS}       % Give a unique label
\end{figure*} 
The resulting LS periodograms for the muon flux and temperature series are shown in Figure \ref{LS}, top and bottom panels, respectively, in the period range between 30 and 7000 days. The three bands represent the power spectrum fluctuations at 1, 2 and $\mathrm{3 ~\sigma}$ of 10000 background time series simulated according to the adopted autoregressive processes of order 10. 
As one can see, in both periodograms the dominant peak, which stands well above $\mathrm{3 ~\sigma}$, corresponds to a $\mathrm{\sim}$ 1-year period. We thus fit the two series with a pure sinusoidal function: the fit is shown as a blue line in Figure \ref{fitseries} and the resulting amplitudes $\mathrm{A}$ and phases $\mathrm{t_0}$ are listed in Table \ref{tabsinus}. 
\begin{table}[b]%The best place to locate the table environment is directly after its first reference in text
\caption{\label{tabsinus}%
Results of the sinusoidal fit, $\mathrm{K+A~cos~\frac{2\pi}{T}(t - t_0)}$, applied to the two time series.\\}
\begin{ruledtabular}
\begin{tabular}{lccdr}
\multicolumn{1}{l}{} & \multicolumn{1}{c}{K [\%]} & \multicolumn{1}{c}{A [\%]} & \multicolumn{1}{c}{$\mathrm{T[days]}$} &\multicolumn{1}{c}{$\mathrm{t_0~[days]}$}
\tabularnewline
\hline\\
\multicolumn{1}{l}{Temperature Series} &\multicolumn{1}{c}{$\mathrm{-0.05\pm0.01}$} & \multicolumn{1}{c}{$\mathrm{1.47\pm0.01}$} &  \multicolumn{1}{c}{$\mathrm{365.1\pm0.1}$} &\multicolumn{1}{c}{$\mathrm{184\pm1}$} 
\tabularnewline
\multicolumn{1}{l}{Muon Series} &\multicolumn{1}{c}{$\mathrm{-0.00\pm0.02}$} & \multicolumn{1}{c}{$\mathrm{1.41\pm0.03}$} &  \multicolumn{1}{c}{$\mathrm{365.1\pm0.2}$} &\multicolumn{1}{c}{$\mathrm{186\pm2}$} 
\tabularnewline
\end{tabular}
\end{ruledtabular}
\end{table}
The amplitudes and phases well-agree with those inferred by other experiments, also in the same underground site \cite{2012Bellini}, \cite{2016Agostini},\cite{2019Agostini}. However, the sinusoidal fit does not describe well either the temperature series ($\mathrm{\chi^2/DoF = 2.8}$) or the muon series  ($\mathrm{\chi^2/DoF = 2.1}$). As other periodicities are noticeable (although with very small significance, above about 1.5 s.d.) in the Lomb-Scargle periodogram, namely at about 0.3 y and 0.5 y for both series, and at about 3 and 10 y for the muon series, we try to better describe both series by including also this sub-leading periodicities. 
To this aim, we apply a Singular Spectrum Analysis (SSA, see \cite{ghil2002asm} and references therein). The SSA uses data-adaptive basis functions instead of sinusoidal ones as for the classical (Fourier) spectral estimates. 
Therefore, it is a very powerful tool to extract amplitudes and frequencies of quasi-periodic components. 
To make the statistical uncertainty smaller than the systematic one, as well as to reduce the computing time for the SSA analysis, we re-bin the two time series into 5-days bins.
The resulting fits are shown as a red line in Figure \ref{fitseries} and the parametrisations are reported in Table \ref{tab:2}, year-by-year, in terms of amplitude and position of minimum and maximum, the latter determined with the accuracy of 2.5 days.
Note that while for the temperature series the amplitudes are quite regular from year to year, they are much less so for the muon series. This difference is most likely due to the combination of the larger fluctuations of the muon data and of the more refined filtering of the SSA smoothing algorithm.    
The reduced chi-squared test when comparing the measured series and the modelled ones, including the sub-leading periodicities yields smaller values than when comparing them to pure sinusoidal models, namely 1.54 and 2.5 for the muon intensity and the temperature, respectively. A specific investigation and possible interpretation of such periodicities goes well beyond the scope of the present work and will be the subject of a successive study exploiting more tailored methods of analysis.

%\begin{landscape}
\begin{table}[htbp]
	%	\centering
	\small
	\hspace{-5.cm}

	\caption{Amplitude, expressed in term of percentage with respect to the total average, and position of minimum and maximum in the $\mathrm{I_{\mu}/I^0_{\mu}}$ and $\mathrm{T_{eff}/T^0_{eff}}$ series, calculated year by year.}
	\vspace{0.5cm}
	\begin{tabular}{|ccc|ccc|c|ccc|ccc|}
		\toprule
		\multicolumn{6}{|c|}{\textbf{MUON FLUX}}      &       & \multicolumn{6}{c|}{\textbf{T\textsubscript{eff}}} \\
		%		\midrule
		\textbf{Day} & \textbf{Date} & \multicolumn{1}{c|}{\textbf{A\textsubscript{max}[\%]}} & \textbf{Day} & \textbf{Date} & \textbf{A\textsubscript{min}[\%]} &       & \textbf{Day} & \textbf{Date} & \multicolumn{1}{c|}{\textbf{A\textsubscript{max}[\%]}} & \textbf{Day} & \textbf{Date} & \textbf{A\textsubscript{min}[\%]} \\
		%		\midrule
	
183   & 02.07.1994 & 0.99  & 358   & 24.12.1994 & -2.99 &       & 203   & 22.07.1994 & 1.56  & 348   & 14.12.1994 & -1.61 \\
563   & 17.07.1995 & 0.86  & 723   & 24.12.1995 & -2.89 &       & 568   & 22.07.1995 & 1.56  & 713   & 14.12.1995 & -1.61 \\
933   & 21.07.1996 & 1.41  & 1088  & 23.12.1996 & -1.54 &       & 933   & 21.07.1996 & 1.56  & 1078  & 13.12.1996 & -1.62 \\
1288  & 11.07.1997 & 1.52  & 1453  & 23.12.1997 & -1.62 &       & 1298  & 21.07.1997 & 1.56  & 1443  & 13.12.1997 & -1.62 \\
1653  & 11.07.1998 & 1.82  & 1828  & 02.01.1999 & -1.89 &       & 1663  & 21.07.1998 & 1.56  & 1808  & 13.12.1998 & -1.62 \\
2028  & 21.07.1999 & 1.66  & 2188  & 28.12.1999 & -1.15 &       & 2028  & 21.07.1999 & 1.57  & 2173  & 13.12.1999 & -1.63 \\
2383  & 10.07.2000 & 1.50  & 2543  & 17.12.2000 & -1.21 &       & 2393  & 20.07.2000 & 1.56  & 2543  & 17.12.2000 & -1.63 \\
2743  & 05.07.2001 & 1.90  & 2918  & 27.12.2001 & -1.77 &       & 2758  & 20.07.2001 & 1.57  & 2908  & 17.12.2001 & -1.63 \\
3128  & 25.07.2002 & 1.70  & 3278  & 22.12.2002 & -1.22 &       & 3123  & 20.07.2002 & 1.56  & 3273  & 17.12.2002 & -1.63 \\
3493  & 25.07.2003 & 1.54  & 3638  & 17.12.2003 & -0.93 &       & 3488  & 20.07.2003 & 1.56  & 3638  & 17.12.2003 & -1.64 \\
3853  & 19.07.2004 & 1.79  & 4003  & 16.12.2004 & -1.17 &       & 3853  & 19.07.2004 & 1.55  & 3998  & 11.12.2004 & -1.62 \\
4213  & 14.07.2005 & 1.31  & 4358  & 06.12.2005 & -1.57 &       & 4218  & 19.07.2005 & 1.56  & 4363  & 11.12.2005 & -1.58 \\
4583  & 19.07.2006 & 1.56  & 4723  & 06.12.2006 & -1.17 &       & 4583  & 19.07.2006 & 1.56  & 4728  & 11.12.2006 & -1.55 \\
4948  & 19.07.2007 & 1.82  & 5098  & 16.12.2007 & -0.94 &       & 4948  & 19.07.2007 & 1.55  & 5093  & 11.12.2007 & -1.50 \\
5308  & 13.07.2008 & 1.50  & 5453  & 05.12.2008 & -1.00 &       & 5313  & 18.07.2008 & 1.55  & 5458  & 10.12.2008 & -1.46 \\
5673  & 13.07.2009 & 1.85  & 5818  & 05.12.2009 & -0.68 &       & 5678  & 18.07.2009 & 1.54  & 5818  & 05.12.2009 & -1.41 \\
6043  & 18.07.2010 & 2.24  & 6183  & 05.12.2010 & -0.10 &       & 6043  & 18.07.2010 & 1.55  & 6183  & 05.12.2010 & -1.39 \\
6398  & 08.07.2011 & 1.88  & 6558  & 15.12.2011 & -1.04 &       & 6408  & 18.07.2011 & 1.57  & 6548  & 05.12.2011 & -1.42 \\
6763  & 07.07.2012 & 1.22  & 6908  & 29.11.2012 & -1.59 &       & 6773  & 17.07.2012 & 1.58  & 6913  & 04.12.2012 & -1.43 \\
7133  & 12.07.2013 & 1.56  & 7268  & 24.11.2013 & -0.71 &       & 7138  & 17.07.2013 & 1.59  & 7273  & 29.11.2013 & -1.45 \\
7488  & 02.07.2014 & 2.15  & 7643  & 04.12.2014 & -1.22 &       & 7503  & 17.07.2014 & 1.61  & 7638  & 29.11.2014 & -1.47 \\
7853  & 02.07.2015 & 1.76  & 8018  & 14.12.2015 & -2.38 &       & 7868  & 17.07.2015 & 1.63  & 8003  & 29.11.2015 & -1.50 \\
8228  & 11.07.2016 & 0.97  & 8378  & 08.12.2016 & -2.50 &       & 8233  & 16.07.2016 & 1.64  & 8373  & 03.12.2016 & -1.52 \\
8603  & 21.07.2017 & 0.77  & 8738  & 03.12.2017 & -1.73 &       & 8598  & 16.07.2017 & 1.62  & 8738  & 03.12.2017 & -1.49 \\

		\toprule
		%		\bottomrule
	\end{tabular}%
	\label{tab:2}%
\end{table}%

%\end{landscape}

\begin{figure*}[htbp]
\centering
\vspace{0.cm}
\includegraphics[width=15.cm]{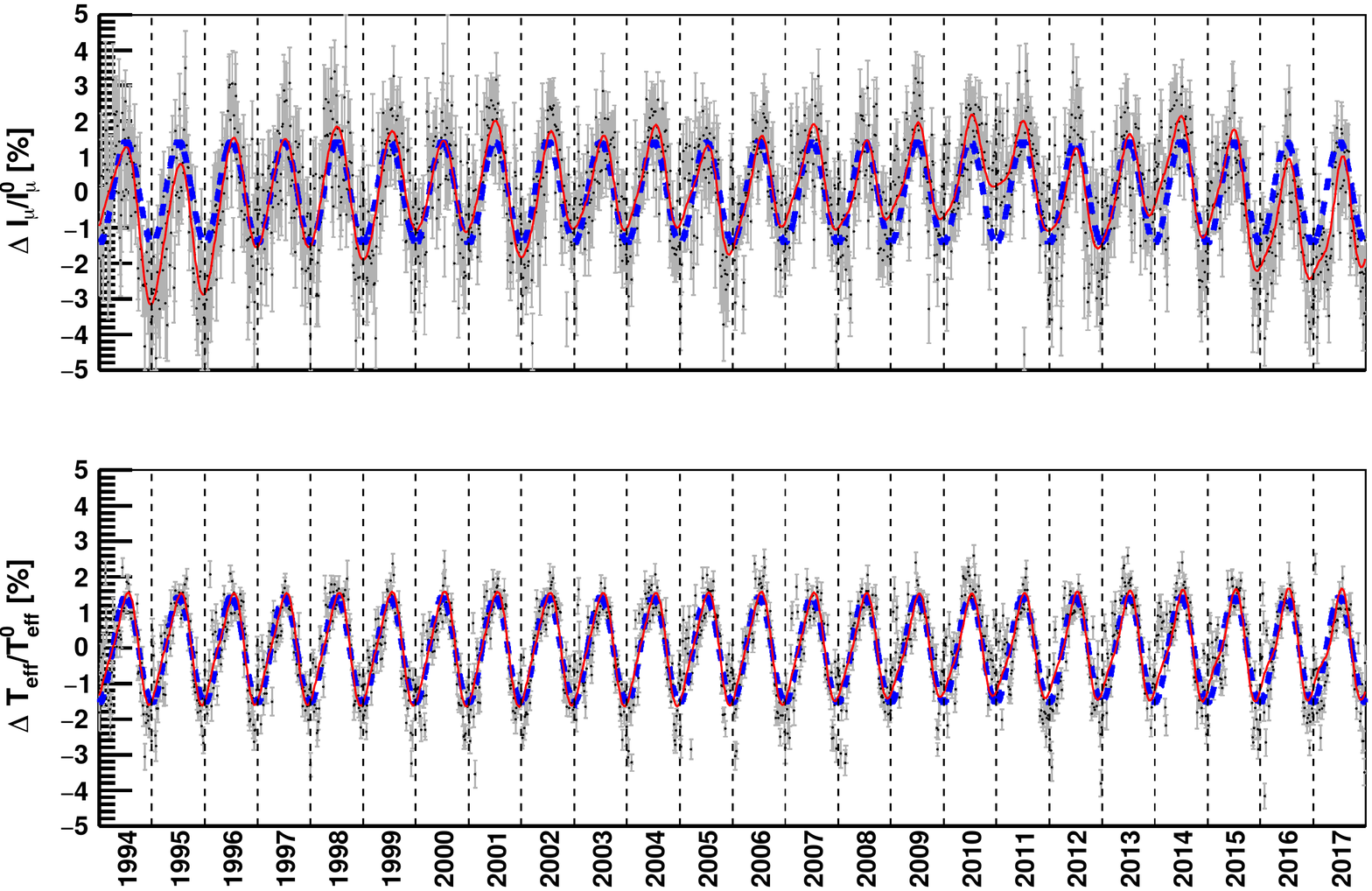}
\vspace{0.0cm}
\caption{{Time series of the muon flux (top panel) and effective temperature (bottom panel) rebinned into 5-days bins. The blue (thick dashed) lines represent the sinusoidal fits of Table \ref{tabsinus}, while the red (thin) lines represent the interpolations obtained via the Singular Spectrum Analysis.}
}
%\vspace{-2.cm}
\label{fitseries}       % Give a unique label
\end{figure*}

\section{Discussion and Conclusions}
\label{sum}

In this paper we have studied the time series of more than fifty millions of muons detected by LVD in 24 years in the Hall A of the LNGS, the longest muon series ever recorded underground. 
We have measured an average muon flux of $(\mathrm{3.35 \pm 0.0005^{stat}\pm 0.03^{sys}) \cdot 10^{-4} ~m^{-2} s^{-1}}$, which is consistent with values previously reported by LVD, as well as with measurements performed in the same laboratory by other experiments.

We have observed that the flux of underground muons is 
%not constant but 
modulated due to the temperature variations in the stratosphere whose main periodicity is seasonal.
We have quantified such a correlation by using the upper-air temperature data set obtained from the European Center for Medium-range Weather Forecasts, finding an effective temperature coefficient, $\mathrm{\alpha_{T}} = 0.94\pm0.01^{stat} \pm0.01^{sys}$.
This measurement is in good agreement with model predictions of muon production from pions and kaons decay as well as with other measurements at the same depth.

The long term monitoring of the muon background is a relevant information for an underground laboratory, especially for long-duration experiments searching for rare events. We have thus investigated the spectral content of the time series of the muon flux by means of the Lomb-Scargle analysis, where we have modeled the random noise with an autoregressive model of order 10. 
The resulting periodogram shows a dominant peak, with a significance much larger than $\mathrm{3~ \sigma}$, corresponding to a period of 1 year. 
We have found indications of additional sub-leading peaks, which support the fact that the series is not a pure sinusoidal wave. By exploiting the SSA analysis, we have characterized the muon series in terms of amplitude and position of maximum and minimum, for the first time on a year-by-year basis.

A specific investigation of such secondary periodicities will be the subject of a dedicated study. 
%Modified ex-referee. Yet, as one of the secondary periodicity corresponds to about 11 years, we comment here in view of an intriguing report on the presence in a sample of Gran Sasso data, including also LVD, of a modulation with a period of about 11 years \cite{the Gran Sasso}. 
Yet, as one of them corresponds to a period of about 10 years, we comment here in view of an intriguing report on the presence in a sample of Gran Sasso data, including also LVD, of a modulation with a period of the same order (about 11 years) \cite{the Gran Sasso}.
The authors of that report found that the power was well above 99\% and that the phase was anti-correlated with the solar cycle.
With the data set used in this work, which is three times larger and where a very accurate study of the noise of the time series has been performed, we have found that the significance associated to the same periodicity is about $\mathrm{1.5~ \sigma}$. In spite of the limited significance, we have evaluated the corresponding phase that is opposite to the one found in \cite{the Gran Sasso}.
We note that a correlation between the stratospheric temperature and the solar cycle has been recently reported for example in \cite{stratos1} and \cite{Randel2009}.

\section{Acknowledgments}
The authors wish to thank all 
%directors and all 
the staff of the National Gran Sasso Laboratory for
their constant support and cooperation during all these years. The successful installation, commissioning, and operation of LVD would not have been possible without the commitment and assistance of the technical staff of all LVD institutions.
Some of the scientists who imagined, realized and contributed to the LVD experiment are not with us anymore. 
We are left with their memory and their teachings.\\

CV gratefully acknowledges the support to this work by the "Departments of Excellence 2018 -2022" Grant awarded by the Italian Ministry of Education, University and Research (MIUR) (L. 232/2016).

%


%Bibliography here

\end{document}